# Taking the Heat Off of Plasmonic Chemistry
*Viewpoint*

*Prashant K. Jain*[1,2,3,4*]

[1]*Department of Chemistry, University of Illinois at Urbana–Champaign, Urbana, IL 61801, United States*
[2]*Materials Research Laboratory, University of Illinois at Urbana–Champaign, Urbana, IL 61801, United States*
[3]*Department of Physics, University of Illinois at Urbana–Champaign, Urbana, IL 61801, Unites States*
[4]*Beckman Institute of Advanced Science and Technology, University of Illinois at Urbana–Champaign, Urbana, IL 61801, United States.*

Corresponding Author E-mail: jain@illinois.edu

**What is Plasmonic Catalysis?** Plasmon-excitation-mediated chemistry, which is a rapidly growing field, is founded on a simple principle: the excitation of the localized surface plasmon resonance (LSPR) of metal nanoparticles triggers chemical reactions on the surfaces of the nanoparticles.[1-4] Early examples of plasmon-excitation-driven nanoparticle synthesis[5,6] and hot-electron-driven chemical reactions induced by ultrashort pulse excitation of metal nanoparticles[7] can be thought of as precursors to the findings of direct photocatalysis by plasmonic nanoparticles[3], which is the focus of this Viewpoint. The field in its current state was, in large part, invigorated by a 2011 paper,[8] which showed that the excitation of the LSPR of Ag nanoparticles by continuous-wave (CW), visible-frequency light triggered the dissociation of adsorbed $O_2$. The $O^.$ atoms thus produced were utilized for industrially relevant oxidation reactions such as those of propylene and ethylene, which would have otherwise required high-temperature and -pressure conditions to proceed at appreciable rates.[8,9] In the absence of visible-light excitation, the bond dissociation and oxidation reactions proceeded at appreciably low rates, which indicated that plasmonic excitation enhanced the rates of these chemical reactions. The phenomenon is, therefore, termed as plasmonic catalysis or plasmonic photocatalysis. Another classic example of plasmonic catalysis involves the dissociation of $H_2$ to $H^.$ on Au nanoparticles under visible-light excitation of the LSPR.[10,11] While a detailed electronic description of these phenomena is still being worked out, one proposed mechanistic picture is that the decay of the LSPR by Landau collisionless damping and electron–electron scattering in the metal results in the intraband (or interband) excitation of an sp-band (or d-band) electron[12] to a state above the equilibrium Fermi level of the metal. The kinetic energy of one or more such hot electrons is transferred to an adsorbate, such as $O_2$, by electron–adsorbate scattering resulting in the nonthermal vibrational excitation of a bond in the adsorbate.[1,8] The excited adsorbate is then much more likely to undergo bond dissociation than one would expect from the equilibrium temperature. It must be noted that for adsorbates that form strong electronic admixtures with states of the metal,[13,14] such electron–adsorbate scattering can take place concomitantly or coherently[1,7,15,16] with decay of the LSPR, a phenomenon termed as chemical interface damping.[13]

**Plasmonic Heating.** Metals by their very electronic nature are lossy,[17] so there is an accompanying set of energy relaxation processes that is invariably involved under plasmonic excitation. Excited electrons in metals undergo relaxation by electron–phonon coupling on the subpicosecond-to-picosecond time scale.[18,19] After a picosecond of the initial excitation, the absorbed energy is no longer carried by excited electrons, but it is shared with the lattice phonons of the nanoparticle. The heated nanoparticle lattice further exchanges energy with the surrounding medium by phonon–phonon coupling to result in a heating of the medium, until the nanoparticle lattice and its surrounding medium reach thermal equilibrium on the 100 ps timescale.[18] As a consequence, under CW light excitation, the nanoparticle surface and the local medium around the nanoparticle are subjected to a steady-state temperature that is higher than the temperature of the bulk medium. Thus, heating of the medium is part-and-parcel of processes involving plasmonic excitation and this photothermal heating effect[20] has been used to good advantage where remote and localized delivery of heat is desired, e.g., photothermal destruction of cancer cells[21] and heat-assisted magnetic recording.[22]

**Hot Electrons or Phonons.** For studies of plasmonic catalysis, photothermal heating poses a complication. The elevated temperature of the nanoparticle surface under CW excitation can itself lead to an enhancement in the rate of a chemical reaction taking place at the surface; after all, the reaction can be considered to be taking place in a higher-temperature bath. A simple application of the Arrhenius law shows that if the steady-state temperature at the nanoparticle surface was increased from 298 K in the dark to 318 K under plasmonic excitation then a chemical reaction, with an apparent activation energy of 1 eV or 96.5 kJ·mol$^{-1}$, would proceed at an order-of-magnitude higher rate than in the absence of plasmonic excitation. In such a case, how does one distinguish the putative action of hot electrons on the reaction rate from the thermal enhancement? From the mechanistic standpoint, the central dilemma can arguably be boiled down to whether the rate enhancement is caused by the action of energetic electrons or by the thermal energy of the heated phonon bath.[23,24] Or, if both are involved, what is the contribution of each?[25,26] These questions have not escaped the attention of experimental researchers in the community.

**The Elusive Control Experiment.** In principle, it is straightforward to measure whether and how much of the rate enhancement is due to the nonthermal action of carriers. If the steady-state temperature of the surface of the nanoparticles under plasmonic excitation, $T_s$, was known from measurements or simulations, then one ought to simply perform a control experiment with no light excitation but with the reactor heated in a manner that the nanoparticle surface temperature is precisely $T_s$. All other conditions should be maintained the same. The rate measured in this control experiment, $R_{dark}$, ought to be the reference point for comparison of the reaction rate under plasmonic excitation, $R$. For instance, if $R/R_{dark} > 1$, then one would conclude that the plasmonic excitation enhanced the reaction rate by way of a nonthermal effect. Although uncommon, rate suppression by light excitation, $R/R_{dark} < 1$, would also suggest a nonthermal effect at play. While such a rate suppression effect would be of no interest to most researchers, it may be exploited in situations where the reaction in question is a side reaction desired to be suppressed.

However, in practice, the control experiment described above is easier said than done. First, it is difficult to precisely measure the surface temperature. One would need a microscopic probe of the surface temperature, e.g., Raman thermometry.[27] Second, such a measurement ought to be carried out under the exact conditions at which the photoreaction rate measurements are performed and preferably should be performed *in situ*. The light excitation attributes, sample configuration, and heat and mass transport characteristics of the medium all influence the surface temperature. A temperature measured under conditions differing from the actual reaction (e.g., on a dry catalyst film, whereas the actual reaction took place on a catalyst immersed in a liquid) is misleading. Heat transport and temperature simulations can offer guidance, provided realistic conditions of the reaction are properly incorporated in the simulation; forced convection or complex geometries can be particularly challenging to simulate. Note the surface temperature estimation, whether performed by measurement or by simulation, should be conducted for the dark control experiment as well; there may be thermal gradients and nonuniformities even in the absence of light excitation, especially when a localized heating source is employed and the medium has poor heat transport characteristics (e.g., a gaseous medium or a viscous liquid with no stirring). The dark reaction rate is a valid reference point only when the surface temperature is the same or nearly the same in the dark control reaction and the plasmonic-excitation-driven reaction. A difference in the surface temperature in the two scenarios will result in a spurious enhancement factor $R/R_0$.

**Recent Debate.** It has been recently argued[23] that in some well-established bond dissociation reactions the rate enhancement attributed to the action of hot electrons is instead solely due to the elevated surface temperature of the nanoparticle catalyst under plasmonic excitation. The group claims that the surface temperature under plasmonic excitation was underestimated, which resulted in the appearance of a nonthermal enhancement where there was none present. The arguments made[28-30] point to the practical challenges of precisely accounting for thermal effects. So, does this imply that there can be no satisfactory resolution of the question about the nonthermal role of plasmonic excitation? Quite the contrary. By following some best practices, discussed in the remainder of the Viewpoint, it is possible to conclusively demonstrate and/or exploit nonthermal effects. In fact, there are published reports that follow these practices and provide unambiguous evidence of a photochemical (nonthermal) role of plasmonic excitation.

**Why Does the Mechanism Matter?** Why, a practitioner may ask, does the mechanistic origin of the enhancement matter as long as the input light energy serves to favorably enhance the production rate of an industrially desired chemical by one or the other means? One could counter that if the mechanism of rate enhancement was purely photothermal then one could instead simply heat the reactor up to achieve a higher temperature, rather than use plasmonic excitation. However, the plasmonic means of heating can be argued to have two potential advantages. First, the heating could be preferably localized to the surface of the nanoparticles, where the adsorbates undergo rate-limiting catalytic reactions. Such localization is invariably achieved when the medium has poor heat conduction and convection properties. Due to its ability to set up a desirable spatial temperature profile, plasmonic heating would be more efficient than bulk-scale heating of the reactor, provided that losses due to plasmonic scattering are minimized. Second, if the infrastructure were available, plasmonic heating could allow the use of an abundant source of visible light, i.e., sunlight, for replacing the energy needs of conventional heating.[31]

On the other hand, if plasmonic enhancement was purely photothermal and no plasmonically generated charge carriers were involved in the chemistry, then there would be little to no prospect of employing plasmonic excitation, as discussed later, for promoting reaction pathways that are not favored under thermal conditions,[32] or for modifying and controlling reaction pathways,[32,33] or, for converting the input light energy into chemical energy stored in fuels.[34] For realistic evaluation of these aspirations, the mechanistic origin of rate enhancement is a critical piece of information.

**Basics of Photothermal Heating.** The fraction of plasmonic excitations that is not radiatively scattered by the nanoparticles undergoes nonradiative decay to heat, which is deposited in the irradiated volume. If the deposited heat is not removed fast enough, the heat will accumulate in the irradiated volume, resulting in a local temperature rise. Under CW illumination, the rate at which heat is deposited depends directly on the absorbed power, $\dot{q}$, which is the product of the incident light intensity, $I$ (W·cm$^{-2}$), and the absorption crosssection, $\sigma$ (cm$^2$):

$$\dot{q} = \sigma I \quad (1)$$

Naturally, decreasing the incident light intensity is not a viable option for minimizing photothermal heating, because any light-assisted or light-driven chemistry would also suffer as a consequence.

Note that for a single plasmonic nanoparticle subject to typical CW illumination intensities (few W·cm$^{-2}$) the rate of heating is small enough that it is easily offset by the rate of cooling and no substantial increase in the temperature of the nanoparticle or its surrounding medium is expected.[35] However, in typical experiments, multiple nanoparticles are present within the irradiated volume, and the rate of heat deposition, which is directly proportional to the volumetric density of the nanoparticles,[35] can be large enough under typical CW illumination intensities to cause a temperature rise.

Even though heating is unavoidable, one can strive to minimize the localization of the heat by using conditions that favor heat transfer out from the nanoparticles into the medium surrounding them. At a fixed rate of heat deposition, the higher the rate of heat transfer, the lower the accumulation of the heat at the nanoparticles and, consequently, the smaller will be the difference between the steady-state temperature of the nanoparticles, $T_s$, and that of the bulk medium, $T_{bulk}$. Note that the entire nanostructure is expected to be at nearly the same temperature, $T_s$, because metals conduct heat effectively, so a thermal hot spot or gradient is unlikely on a nanometer-length scale under typical CW light intensities.

Heat propagates away from the nanostructure by its conduction into the surrounding medium, by convection (both free and forced), and by radiation. A steady state is attained when the rate of heat deposition, $\dot{q}$, is balanced by the rate of heat removal. This balance is sustained by temperature difference between the irradiated nanoparticles and the bulk medium:[36]

$$\dot{q} = \kappa A \frac{dT}{dx} + hA(T_s - T_{bulk}) \quad (2)$$

where $\kappa$ is the thermal conductivity of the medium; $\frac{dT}{dx}$ is the temperature gradient; $h$ is the heat transfer coefficient for convection; $A$ is the area normal to the direction of heat transfer; and heat radiation is neglected since it has a relatively small contribution unless the temperature is 100s of K or higher. From eq 2, it is seen that the use of a medium with a higher thermal conductivity $\kappa$ would result in a smaller temperature difference between the nanoparticle surface and the bulk medium. In this regard, liquid media are typically more favorable than gases. Liquid water has a room-temperature $\kappa$ of 0.6 W·m$^{-1}$·K$^{-1}$,[36] 25× higher than that of water vapor; therefore, in liquid water, the temperature difference, $\Delta T = T_s - T_{bulk}$, would be, to a first approximation, 25× smaller, given all other conditions are the same. Stirring or agitation of the liquid or gaseous medium allows forced convection of the heat, which is considerably more effective for prevention of heat localization as compared to natural convection resulting from density differences in the fluid. For instance, the heat transfer coefficient, $h$, of water is estimated to be 100–1,000 W·m$^{-2}$·K$^{-1}$ for free convection, whereas for forced convection, it can range from 500 to 10,000 W·m$^{-2}$·K$^{-1}$, the precise value[37] depending on flow properties such as the mass flow rate and degree of turbulence, the temperature, and possibly also on the nature of the nanoscale interface. From the point of view of convection, liquids are typically superior to gases. A vigorously stirred liquid medium is ideal from the standpoint of minimization of localized heating of the nanoparticle.

**Colloidal Photoreactions.** The factors discussed above lead us to a practical experimental system where photothermal heating effects can be accounted for most effectively: a photoreaction performed in a colloid of plasmonic nanoparticles, irradiated by CW light.[33,38-40] In a well-dispersed state, contact between individual nanoparticles and the liquid medium ensures effective heat removal by conduction. Stirring of the colloid further enhances heat and mass transport, ensuring homogenization of the heat even if the CW light is focused on a portion of the reactor. In such a system, the difference between the nanoparticle surface temperature and the bulk temperature:[35]

$$\Delta T = T_s - T_{bulk} = \frac{\sigma I}{4\pi \kappa r} \quad (3)$$

can be reduced to an insubstantial magnitude at typical CW illumination intensities. For instance, for a nanoparticle radius, $r$, of 6 nm, a light intensity as high as 1 kW.cm$^{-2}$, an absorption crosssection of 50 nm$^2$ (typical for Au nanoparticles of 6 nm radius at their LSPR wavelength maximum)[33] and water as the liquid medium, $\Delta T$ is ~10$^{-2}$ K. In such a case, the bulk medium temperature, which can be measured more directly or easily, becomes a proxy of the nanoparticle surface temperature. Since $T_s$ becomes effectively known, the control experiment, described in a previous section, is no longer elusive and can be appropriately performed, as exploited in some model systems.[33,38]

It must be acknowledged that colloidal photoreactions are impractical for the study of industrially relevant chemical reactions and/or for accomplishing large scales. For such objectives, it is fitting to employ reactors where gas-phase reactants are flowed through a packed bed consisting of solid pellets of the plasmonic catalyst. Such systems, however, present complex heat transport characteristics and temperature distributions, making it challenging to precisely account for or systematically control photothermal heating effects and requiring care.[41] Therefore, advances enabled by these latter types of reactions should be complemented by systematic mechanistic investigations made possible by colloidal photoreactions.

**Expanding the Space of Chemical Reactions**. While appropriately designed experiments enable us to account for the photothermal effect of plasmonic excitation and thereby resolve nonthermal effects, the question that remains is are we fully exploiting the potential action of plasmonic excitation? The answer lies in the choice of chemical processes we study. It has been common for practitioners to demonstrate or utilize plasmonic excitation as a means, alternative to the use of high temperatures, for enhancing the kinetics of a chemical reaction. At the origin of such plasmonic enhancement is thought to be an increase in the adsorbate concentration,[28] the access of an excited-state potential energy surface,[28] or a decrease in the activation barrier brought about by the energetic carriers produced by plasmonic excitation.[38] This strategy may have practical advantages, for instance, where high-temperature operation causes restructuring or deactivation of the nanoparticulate catalyst. However, the aspirations of the community ought to expand beyond the *plasmonic enhancement* of chemical processes that are already known to take place, albeit at small rates. We should increasingly strive to discover *plasmon-driven chemistries* that are simply not possible in the absence of light excitation, chemistries where the plasmon serves as an indispensable "chemical reagent". Some examples of such avenues follow:

i) Chemical reactions where plasmonic excitation switches the reaction to a different pathway than the one followed in the conventional thermal reaction. In other words, the plasmonic excitation is used to direct or switch reaction selectivity.[32,33]

ii) Reactions that involve electron transfer steps (i.e., redox reactions)[33,38-40,42] and do not proceed unless electron–hole pairs generated by the damping of the plasmonic excitation are utilized. In such cases, it would not be sufficient for the energy of hot carriers to be harvested, but both charge carriers themselves would need to be extracted by acceptor–donor pairs.

iii) Processes that involve chemical bond formation, rather than bond dissociation in the surface-catalyzed step.

iv) Reactions that have no measurable rate in the dark,[34] even at elevated temperatures, unless plasmonic excitation is employed. Of course, the concept of a "zero rate" is somewhat subjective since it simply means that the generation of products, if any, is below some practical detection limit. In terms of a thermochemical criterion, which is more objective, reactions which are thermodynamically nonspontaneous (free energy of reaction, $\Delta G > 0$) in the forward direction and therefore require a source of Gibb's free energy would fall under this category.

None of these chemical objectives would be achievable simply by thermal means. The nonthermal chemical action of plasmonic excitation would undoubtedly need to be exploited.

**Plasmonic Photocatalysis vs. Plasmonic Photosynthesis.** Based on a textbook-level thermochemical classification of chemical reactions (Table 1), one can envisage two distinct categories for plasmonic chemistry:

i) **Plasmonic photocatalysis**, which ought to refer to the use of plasmonic excitation of the catalyst for enhancement of the kinetics of an exergonic reaction ($\Delta G < 0$). Such a reaction is spontaneous under the given conditions even without light excitation; plasmonic excitation assists the reaction by enhancing its rate. The rate enhancement can have a thermal contribution, which must be carefully accounted for, as described in the previous sections, so that nonthermal enhancement of the reaction kinetics can be properly assigned.

ii) An endergonic or endoergic reaction ($\Delta G > 0$) is nonspontaneous under the given conditions. In the absence of an external source of free energy, the reaction does not advance in the forward direction. There is no rate, to speak of, that can be enhanced, whether thermally or nonthermally, unless free energy is provided. In such a case, the plasmon excitation scheme would need to be capable of utilizing the free energy of light, typically via the formation of energy-rich intermediates, to perform chemical work, i.e., drive the uphill reaction. Essentially, such a scheme would constitute the use of light energy for the synthesis of a chemical fuel, which is why it would be appropriate to categorize such a scheme as **plasmonic photosynthesis**. In such a scheme, the plasmonic excitation plays a role distinct from a kinetic one by providing free energy indispensable for driving the reaction. Any accompanying reduction in the reaction activation barrier and/or photothermal enhancement of the reaction rate constitute only ancillary effects. Plasmonic photosynthesis naturally encompasses plasmonic photocatalysis but goes well beyond it.

**Table 1.** Classification of Chemical Reactions by Thermochemistry; Temperature is Denoted as $T$.

| $\Delta H$ | $\Delta S$ | $\Delta G$ | outcome | type of plasmonic chemistry |
|---|---|---|---|---|
| -ve | +ve | -ve | spontaneous at any $T$ | plasmonic photocatalysis |
| +ve | +ve | +ve or -ve | nonspontaneous at low $T$ spontaneous at high $T$ | depends on $T$ |
| -ve | -ve | -ve or +ve | spontaneous at low $T$ nonspontaneous at high $T$ | depends on $T$ |
| +ve | -ve | +ve | nonspontaneous at any $T$ | plasmonic photosynthesis |

As an illustration of the categorization described above, let us consider two common reactions. The oxidation of $H_2$, $2H_2 + O_2 \rightarrow 2H_2O$, would not require a free energy input under standard conditions but would benefit from plasmonic photocatalysis (or thermocatalysis) of the otherwise sluggish $H_2$ and $O_2$ dissociation steps to form reactive $H^.$ and $O^.$. On the other hand, the reduction of $CO_2$ accompanied by the oxidation of water, $CO_2 + 2H_2O \rightarrow CH_4 + 2O_2$, involves, under standard conditions, the thermodynamically uphill formation of a chemical fuel $CH_4$ and would therefore necessitate chemical work to be performed by plasmonic excitation.

There are some complications in this simplified classification that must be mentioned. First, the treatment assumes a closed system, which is often not the case. Second, it is not sufficient to use standard free energies, enthalpies, or entropies of reactions; corrections for nonstandard conditions, especially the concentrations (activities) or pressures (fugacities) of all reactants and products involved in the reaction, are necessary. Third, the classification does not take into account processes involving a phase change (e.g., solar water vapor generation). As seen in Table 1, for reactions with a –ve $\Delta H$ and a –ve $\Delta S$, neglecting any temperature dependence of the thermodynamic parameters themselves, a large upswing in the temperature can cause a transition from a spontaneous regime to a nonspontaneous one. Conversely, for reactions with a +ve $\Delta H$ and a +ve $\Delta S$, a large upswing in the temperature can cause a transition from a nonspontaneous regime to a spontaneous one. In such cases, the simple categorization does not apply. A large enough temperature upswing, if caused by plasmonic heating, can entropically "suppress" or "advance" the forward reaction. However, such large temperature changes would be precluded by following the best practices described in the previous discussion.

The final row in Table 1 deserves special mention. These reactions involve a +ve $\Delta H$ and -ve $\Delta S$ and are nonspontaneous at any temperature. Driving such reactions constitutes an ultimate test of the nonthermal, photochemical action of plasmon excitation. Natural photosynthesis, involving the conversion of $CO_2$ and $H_2O$ to glucose and $O_2$, involves both a +ve $\Delta H$ and -ve $\Delta S$.[43] Note that the net entropy loss in the photosynthetic system is more than offset by an increase in the entropy of solar radiation as it transitions from a lower entropy state (higher Planck temperature of the radiation source) to a higher entropy state (lower Planck temperature of the absorber in equilibrium with the heat bath comprised by the reaction mixture).[44] Effectively, the internal energy of light quanta and the low entropy of solar radiation are utilized for photosynthesis. Plasmonic photosynthesis represents an analogous opportunity and challenge to harvest the Gibb's free energy of visible light.

**Acknowledgements** This material is based upon work supported by the National Science Foundation under Grant NSF CHE-1455011. P.K.J. thanks Varun Mohan and Sungju Yu for proof-reading a draft version. The author has no competing financial interests.

**References**

1. Linic, S.; Aslam, U.; Boerigter, C.; Morabito, M. Photochemical Transformations on Plasmonic Metal Nanoparticles. *Nat Mater.* **2015**, *14,* 567–576.

2. Hou, W.; Cronin, S. B. A Review of Surface Plasmon Resonance-Enhanced Photocatalysis. *Adv. Funct. Mater.* **2013**, *23*, 1612–1619.

3. Kale, M.J.; Avanesian, T.; Christopher, P. Direct Photocatalysis by Plasmonic Nanostructures. *ACS Catal.* **2014**, *4*, 116–128.

4. Smith, J. G.; Faucheaux, J. A.; Jain, P. K. Plasmon Resonances for Solar Energy Harvesting: A Mechanistic Outlook. *Nano Today* **2015**, *10*, 67–80.

5. Jin, R; Cao, Y; Mirkin, C. A.; Kelly, K. L.; Schatz, G. C.; Zheng, J. G. Photoinduced Conversion of Silver Nanospheres to Nanoprisms. *Science.* **2001**, *294*, 1901–1903.

6. Maillard, M; Huang, P.; Brus, L. Silver Nanodisk Growth by Surface Plasmon Enhanced Photoreduction of Adsorbed [$Ag^+$]. *Nano Lett.* **2003**, *311*, 1611–1615.

7. Jain, P.K.; Qian, W.; El-Sayed, M. A. Ultrafast Cooling of Photoexcited Electrons in Gold Nanoparticle-Thiolated DNA Conjugates Involves the Dissociation of the Gold−Thiol Bond. *J. Am. Chem. Soc.* **2006**, *128*, 2426–2433.

8. Christopher, P.; Xin, H.; Linic, S. Visible-Light-Enhanced Catalytic Oxidation Reactions on Plasmonic Silver Nanostructures. *Nat. Chem.* **2011**, *3*, 467–472.

9. Christopher, P.; Xin, H.; Marimuthu, A.; Linic, S. Singular Characteristics and Unique Chemical Bond Activation Mechanisms of Photocatalytic Reactions on Plasmonic Nanostructures. *Nat. Mater.* **2012**, *11*, 1044–1050.

10. Mukherjee, S.; Libisch, F.; Large, N.; Neumann, O.; Brown, L. V.; Cheng, J.; Lassiter, J. B.; Carter, E. A.; Nordlander, P.; Halas, N. J. Hot Electrons Do the Impossible: Plasmon-Induced Dissociation of $H_2$ on Au. *Nano Lett.* **2013**, *13*, 240–247.

11. Mukherjee, S.; Zhou, L.; Goodman, A. M.; Large, N.; Ayala-Orozco, C.; Zhang, Y.; Nordlander, P.; Halas, N. J. Hot-Electron-Induced Dissociation of $H_2$ on Gold Nanoparticles Supported on $SiO_2$. *J. Am. Chem. Soc.* **2014**, *136*, 64–67.

12. Sundararaman, R.; Narang, P.; Jermyn, A. S.; Goddard III, W. A.; Atwater, H. A. Theoretical Predictions for Hot-Carrier Generation from Surface Plasmon Decay. *Nat. Commun.* **2014**, *5*, 5788.

13. Persson, B. N. J. Polarizability of Small Spherical Metal Particles: Influence of The Matrix Environment. *Surf. Sci.* **1993**, *281*, 153–162.

14. Kumari, G.; Zhang, X.; Devasia, D.; Heo, J.; Jain, P. K. Watching Visible Light-Driven $CO_2$ Reduction on a Plasmonic Nanoparticle Catalyst. *ACS Nano* **2018**, *12*, 8330–8340.

15. Foerster, B.; Spata, V. A.; Carter, E. A.; Sönnichsen, C.; Link, S. Plasmon Damping Depends on The Chemical Nature of The Nanoparticle Interface, *Sci. Adv.* **2019**, *5*, eaav0704.

16. Seemala, B.; Therrien, A. J.; Lou, M.; Li, K.; Finzel, J. P.; Qi, J.; Nordlander, P.; Christopher, P. Plasmon-Mediated Catalytic $O_2$ Dissociation on Ag Nanostructures: Hot Electrons or Near Fields? *ACS Energy Lett.* **2019**, *4*, 1803–1809.

17. Ndukaife, J. C.; Shalaev, V. M.; Boltasseva, A. Plasmonics—Turning Loss into Gain. *Science*, **2016**, *351*, 334–335.

18. Link, S.; El-Sayed, M. A. Spectral Properties and Relaxation Dynamics of Surface Plasmon Electronic Oscillations in Gold and Silver Nanodots and Nanorods. *J. Phys. Chem. B* **1999**, *103*, 8410–8426.

19. Hodak, J. H.; Martini, I.; Hartland, G. V. Spectroscopy and Dynamics of Nanometer-Sized Noble Metal Particles. *J. Phys. Chem. B* **1998**, *102*, 6958–6967.

20. Govorov, A. O.; Richardson, H. H. Generating Heat With Metal Nanoparticles. *Nano Today* **2007**, 2, 30–38.

21. Hirsch, L. R.; Stafford, R. J.; Bankson, J. A.; Sershen, S. R.; Rivera, B.; Price, R. E.; Hazle, J. D.; Halas, N. J.; West, J. L. Nanoshell-Mediated Near-Infrared Thermal Therapy of Tumors Under Magnetic Resonance Guidance. *Proc. Nat. Acad. Sci.* **2003**, *100*, 13549–13554

22. Challener, W. A.; Peng, C.; Itagi, A. V.; Karns, D.; Peng, W.; Peng, Y.; Yang, X.; Zhu, X.; Gokemeijer, N. J.; Hsia, Y.-T.; Ju, G.; Rottmayer, R. E.; Seigler, M. A.; Gage, E. C. Heat-Assisted Magnetic Recording by a Near-Field Transducer with Efficient Optical Energy Transfer. *Nat. Photonics* **2009**, *3*, 220–224.

23. Sivan, Y.; Un, I. W.; Dubi, Y. Thermal Effects–An Alternative Mechanism for Plasmonic-Assisted Photo-Catalysis. *arXiv preprint,* **2019,** arXiv:1902.03169 [physics.chem-ph].

24. Jain, P.K. Phenomenological Arrhenius Analyses in Plasmon-Enhanced Catalysis. *arXiv preprint,* **2019,** arXiv:1908.05373 [physics.chem-ph].

25. Zhang, X.; Li, X.; Reish, M. E.; Zhang, D.; Su, N. Q.; Gutiérrez, Y.; Moreno, F.; Yang, W.; Everitt, H. O.; Liu, J. Plasmon-Enhanced Catalysis: Distinguishing Thermal and Nonthermal Effects. *Nano Lett.* **2018,** *183*, 1714–1723.

26. Yu, Y.; Sundaresan, V.; Willets, K. A. Hot Carriers versus Thermal Effects: Resolving the Enhancement Mechanisms for Plasmon-Mediated Photoelectrochemical Reactions. *J. Phys. Chem. C* **2018**, *122*, 5040–5048.

27. Sarhan, R. M.; Koopman, W.; Schuetz, R.; Schmid, T.; Liebig, F.; Koetz, J.; Bargheer, M. The Importance of Plasmonic Heating for the Plasmon-Driven Photodimerization of 4-Nitrothiophenol. *Sci. Rep.* **2019**, 9, 3060.

28. Zhou, L.; Swearer, D. F.; Zhang, C.; Robatjazi, H.; Zhao, H.; Henderson, L.; Dong, L.; Christopher, P.; Carter, E. A.; Nordlander, P.; Halas, N. J. Quantifying Hot Carrier and Thermal Contributions in Plasmonic Photocatalysis. *Science* **2018**, *362*, 69–72.

29. Sivan, Y.; Baraban, J.; Un, I. W.; Dubi, Y. Comment on "Quantifying hot carrier and thermal contributions in plasmonic photocatalysis" *Science* **2019**, *364*, eaaw9367.

30. Zhou, L.; Swearer, D. F.; Robatjazi, H.; Alabastri, A.; Christopher, P.; Carter, E. A.; Nordlander, P.; Halas, N. J. Response on Comment to "Quantifying Hot Carrier and Thermal Contributions in Plasmonic Photocatalysis" *Science* **2019**, *364*, 69–72.

31. Neumann, O.; Urban, A. S.; Day, J.; Lal, S.; Nordlander, P. J.; Halas, N. J. Solar Vapor Generation Enabled by Nanoparticles. *ACS Nano*, **2013**, *7*, 42–49.

32. Zhang, X.; Li, X.; Zhang, D.; Su, N. Q.; Yang, W.; Everitt, H. O.; Liu, J. Product Selectivity in Plasmonic Photocatalysis for Carbon Dioxide Hydrogenation. *Nat. Commun.* **2017**, *8*, 14542.

33. Yu. S.; Wilson, A. J.; Heo, J.; Jain, P.K. Plasmonic Control of Multi-Electron Transfer and C–C Coupling in Visible-Light-Driven $CO_2$ Reduction on Au Nanoparticles. *Nano Lett.* **2018**, *18*, 2189–2194.

34. Yu. S.; Jain, P.K. Plasmonic Photosynthesis of $C_1$–$C_3$ Hydrocarbons from Carbon Dioxide Assisted by an Ionic Liquid. *Nat. Commun.* **2019**, *10*, 2022.

35. Keblinski, P.; Cahill, D. G.; Bodapati, A.; Sullivan, C. R.; Taton, T. A. Limits of Localized Heating by Electromagnetically Excited Nanoparticles. *J. Appl. Phys.* **2006**, *100*, 054305.

36. Incropera, F. P. and de Witt, D. O. *Fundamentals of Heat and Mass Transfer;* John Wiley and Sons Inc., Hoboken, NJ, 1990.

37. Shah, Y. T. *Thermal Energy: Sources, Recovery, and Applications,* CRC Press, Boca Raton, FL, 2018.

38. Kim, Y.; Dumett Torres, D.; Jain, P. K. Activation Energies of Plasmonic Catalysts. *Nano Lett.* **2016**, *16*, 3399–3407.

39. Kim, Y.; Wilson, A. J.; Jain, P. K. The Nature of Plasmonically Assisted Hot-Electron Transfer in a Donor-Bridge-Acceptor Complex. *ACS Catal.* **2017**, *7*, 4360–4365.

40. Kim, Y.; Smith, J. G.; Jain, P. K. Harvesting Multiple Electron–Hole Pairs Generated Through Plasmonic Excitation of Au Nanoparticles. *Nat. Chem.* **2018**, *10*, 763–769.

41. Li, X.; Zhang, X.; Everitt, H. O.; Liu, J. Light-Induced Thermal Gradients in Ruthenium Catalysts Significantly Enhance Ammonia Production. *Nano Lett.* **2019**, *193*, 1706–1711.

42. Zhai, Y. et al. Polyvinylpyrrolidone-Induced Anisotropic Growth of Gold Nanoprisms in Plasmon-Driven Synthesis. *Nat. Mater.* **2016**, *15*, 889–895.

43. Brittin, W.; Gamow, G. Negative Entropy and Photosynthesis. *Proc. Nat. Acad. Sci.* **1961**, *59*, 724–727.

44. Lavergne, J. Commentary on: "Photosynthesis and negative entropy production" by Jennings and coworkers. *Biochim. Biophys. Acta. Bioenerg.* **2006**, *1757*, 1453–1459.